\documentclass[10pt,oneside,onecolumn,preprint,number,sort,compress]{elsarticle}

\usepackage{amsmath}
\usepackage{amssymb}
\usepackage{subfigure}
\usepackage{hyperref}
\usepackage{mathrsfs}
\usepackage{amsxtra}
\usepackage{mathrsfs}
\usepackage{amsfonts}
\usepackage{dsfont}
\usepackage{dcolumn}
\usepackage{bm}
\usepackage{graphicx}
\usepackage{epstopdf}
\usepackage{txfonts}
\usepackage{braket}
\usepackage{comment}
\usepackage{blkarray}

\usepackage{tikz} 
\usetikzlibrary{shapes.geometric, arrows} 
\tikzstyle{arrow} = [thick,->,>=stealth] 
\tikzset{mynode/.style={inner sep=2pt,fill,outer sep=0,circle}}

\begin{document}

\begin{frontmatter}
\title{QCD Analysis of $\Delta S=0$ Hadronic Parity Violation}
\author[val]{Susan Gardner}\ead{gardner@pa.uky.edu}
\author[val]{Girish Muralidhara}\ead{girish.muralidhara@uky.edu}
\address{Department of Physics and Astronomy, University of Kentucky,
Lexington, Kentucky 40506-0055 USA}

\begin{abstract}
We present a QCD analysis of the effective weak Hamiltonian at hadronic energy scales for strangeness-nonchanging ($\Delta S=0$) hadronic processes. 
Performing a leading-order renormalization group analysis in QCD from the $W$ to the ${\cal O}(2\,\rm GeV)$ energy scale, we derive the pertinent effective Hamiltonian for hadronic parity violation, including the effects of both neutral and charged weak currents.
We compute the complete renormalization group evolution of all isosectors and the evolution through heavy-flavor thresholds for the
first time. We show that 
the additional four-quark operators that enter below the $W$ mass scale from QCD operator mixing effects form a closed set, and they result in a 12×12 anomalous dimension matrix. Computing the resulting effective Hamiltonian and
comparing to earlier results, we affirm the importance of operator mixing effects 
and find, as an example, that the 
parity-violating pion-nucleon coupling constant, 
using the factorization {\it Ansatz} and an assessment
of the pertinent quark charge of the nucleon in 
lattice QCD at the 2 GeV scale, 
is in better agreement with recent experiments. 
\end{abstract}
\end{frontmatter}

\section{Introduction}
\label{sec:Introduction}
In the Standard Model (SM), the observed failure of parity conservation in 
the low-energy interactions of nucleons and nuclei follows
from a subtle interplay of electroweak and strong interaction effects, 
with the nonperturbative nature of the strong interaction acting to confound 
the theoretical interpretation of the effects 
observed thus far. The low-energy nature of these studies 
has meant that 
its theoretical description has focused on phenomenological realizations in hadronic degrees of
freedom, with the long-held expectation that the charged pion exchange
interaction should strongly dominate~\cite{Desplanques:1979hn}. 
Despite the small mass of the pion relative to the $1\, \rm GeV$ scale, 
this has not proven to be the case; 
rather, isoscalar and isotensor interactions, 
also appear to play important phenomenological 
roles~\cite{Phillips:2014kna,Schindler:2015nga,Gardner2017paradigm,NPDGamma:2018vhh,n3He:2020zwd}. 

Direct theoretical insight on 
the relative importance of isoscalar, 
isotensor, and isovector parity-violating nucleon-nucleon (NN) interactions has
come from the analysis of NN amplitudes 
in pionless effective field theory (EFT)
in the large number of colors ($N_c$) limit~\cite{Zhu:2009,Phillips:2014kna,Schindler:2015nga}.
In this paper we revisit this issue 
within a framework that makes direct contact to the degrees
of freedom of the 
Standard Model (SM) Lagrangian. That is, starting with the
effective, flavor-conserving, parity-violating Hamiltonian of quarks apropos to NN 
interactions at the weak scale, 
we use renormalization group techniques, including leading-order (LO) 
QCD evolution and operator mixing effects, matching across heavy-quark thresholds, 
to determine the effective weak Hamiltonian for $u,d$, and $s$ quarks at the $2\,\rm GeV$ scale. 
Work of this kind exists in the literature, starting with that of Ref.~\cite{Desplanques:1979hn},
though the existing work has either made additional calculational 
approximations~\cite{Desplanques:1979hn,Dubovik:1986pj,Tiburzi:2012xx} or has 
specialized to the isovector case~\cite{Dai:1991bx,kaplan1993analysis,Tiburzi:2012hx}. 
In this work we consider all three isosectors, and since 
the 
renormalization group effects we consider 
respect isospin symmetry, 
we separate the problem as 
illustrated
in Fig.~\ref{Fig:rectangle} --- we can determine the isospin-separated effective Hamiltonian
just below the $W$ mass scale and evolve it to hadronic scales {\it or} we can evolve the full
effective Hamiltonian and effect the isospin separation at the same low-energy scale and find the same result. 
As an application, we use our 
weak effective Hamiltonian we have constructed to compute
the parity-violating pion-NN coupling constant that appears in the parity-violating 
Hamiltonian of Desplanques, Donoghue, and Holstein (DDH)~\cite{Desplanques:1979hn}, 
to find improved agreement with 
experimental results. 

\begin{figure}
\centering
\begin{tikzpicture}[scale=3,line width=1pt]
\node (1) {$\vec{C}(M_W)$};
\node (2) [ right of=1, xshift=2cm] {$\vec{C}^{I=1}(M_W)$};
\node (4) [below of=1] {$\vec{C}(\Lambda)$};
\node (3) [ right of=4, xshift=2cm] {$\vec{C}^{I=1}(\Lambda$)};
\draw [arrow] (1) --node[anchor=east] {RG} (4);
\draw [arrow] (1) --node[anchor= south] {I=1 extract} (2);
\draw [arrow] (2) --node[anchor=west] {RG} (3);
\draw [arrow] (4) --node[anchor= north] {I=1 extract} (3); 
{(\textit{h\,k\,l})};
\end{tikzpicture}
\hspace{1.5cm}
\begin{tikzpicture}[scale=3,line width=1pt]
\node (1) {$\vec{C}(M_W)$};
\node (2) [right of=1, xshift=3cm] {$\vec{C}^{I=0\oplus 2}(M_W)$};
\node (4) [below of=1] {$\vec{C}(\Lambda)$};
\node (3) [right of=4, xshift=3cm] {$\vec{C}^{I=0\oplus2}(\Lambda$)};
\draw [arrow] (1) --node[anchor=east] {RG} (4);
\draw [arrow] (1) --node[anchor= south] {I=0$\oplus$2 extract} (2);
\draw [arrow] (2) --node[anchor=west] {RG} (3);
\draw [arrow] (4) --node[anchor= north] {I=0$\oplus$2 extract} (3); 
{(\textit{h\,k\,l})};
\end{tikzpicture}
\centering
\caption{Illustration of renormalization group flow in 
LO QCD with and without isospin separation.}
\label{Fig:rectangle}
\end{figure}

\section{Effective Hamiltonian framework}
We start by building 
an effective theory at the $W$ mass scale, 
comprised of five open flavors of quarks, and then 
we use QCD renormalization group (RG) techniques
to evolve it to 
hadronic energy scales, $\Lambda \sim 2\, \rm GeV$, for which
only the three dynamical quarks, u, d, and s are pertinent.
Thus we begin by considering just these three flavors. 
Summing the contributions from all the $\Delta S=0$ tree-level diagrams, we get the Hamiltonian at the $W$ mass scale. 
We extract 
the parity violating (PV) parts from each amplitude to form the PV Hamiltonian,
keeping the charged- and neutral-gauge boson exchange sectors separate:
\begin{equation}\label{PVH}
    \mathcal{H}^{\rm PV}_{\rm eff} = \mathcal{H}^{\rm PV}_Z + \mathcal{H}^{\rm PV}_W \,,
\end{equation}
where for the $Z^0$ sector 
\begin{equation}
\begin{split}
   &\mathcal{H}^{\rm PV}_Z(M_W) = \frac{G_F s_w^2}{3\sqrt{2}} \left( \Theta_1  - 3( \frac{1}{2s_w^2}-1) \Theta_5\right)\\
    \Theta_1 &= [(\bar{u}u)_V+(\bar{d}d)_V+(\bar{s}s)_V]^{\alpha\alpha}[(\bar{u}u)_A-(\bar{d}d)_A-(\bar{s}s)_A]^{\beta\beta}\\
    \Theta_5 &= [(\bar{u}u)_V-(\bar{d}d)_V-(\bar{s}s)_V]^{\alpha\alpha}[(\bar{u}u)_A-(\bar{d}d)_A-(\bar{s}s)_A]^{\beta\beta}\,. \\
    \end{split}
\end{equation}
Note, e.g., that $(\bar{u}{u})_V^{\alpha\alpha} (\bar{d}{d})_A^{\beta\beta}\equiv 
(\bar{u}^\alpha\gamma^\mu {u}^\alpha) (\bar{d}^\beta \gamma_{\mu}\gamma_5 {d}^\beta)$, where $\alpha$ and $\beta$ 
are color indices, with 
our enumeration of the 
different 4-quark operators 
anticipating later 
developments. 
For the $\Delta S=0 \, W^\pm$ sector, we include the pertinent 
Cabibbo angle 
contributions:
\begin{equation}
\begin{split}
    \mathcal{H}^{\rm PV}_W(M_W) &= \frac{G_F s_w^2}{3\sqrt{2}} \left(\frac{-3}{s_w^2} (\cos^2 \!\theta_c)\Theta_9+\frac{-3}{s_w^2}(\sin^2 \theta_c)
    \Theta_{11} \right)\\ 
    \Theta_9 &= (\bar{u}d)_V^{\alpha \alpha}(\bar{d}u)_A^{\beta \beta}+(\bar{d}u)_V^{\alpha \alpha}(\bar{u}d)_A^{\beta \beta}\\
    \Theta_{11} &= (\bar{u}s)_V^{\alpha \alpha}(\bar{s}u)_A^{\beta \beta}+(\bar{s}u)_V^{\alpha \alpha}(\bar{u}s)_A^{\beta \beta} \,
\end{split}
\end{equation}
with $\lambda \equiv \sin \theta_c=0.2253$, so that
our expression is accurate to ${\cal O}(\lambda^4)$.
Moreover, 
$s_w^2=0.231$, and 
$G_F=1.166 \times 10^{-5}\, \rm{GeV^{-2}}$~\cite{Zyla:2020zbs}. 

\begin{figure}[!h]
    \centering
    \includegraphics[width=9cm]{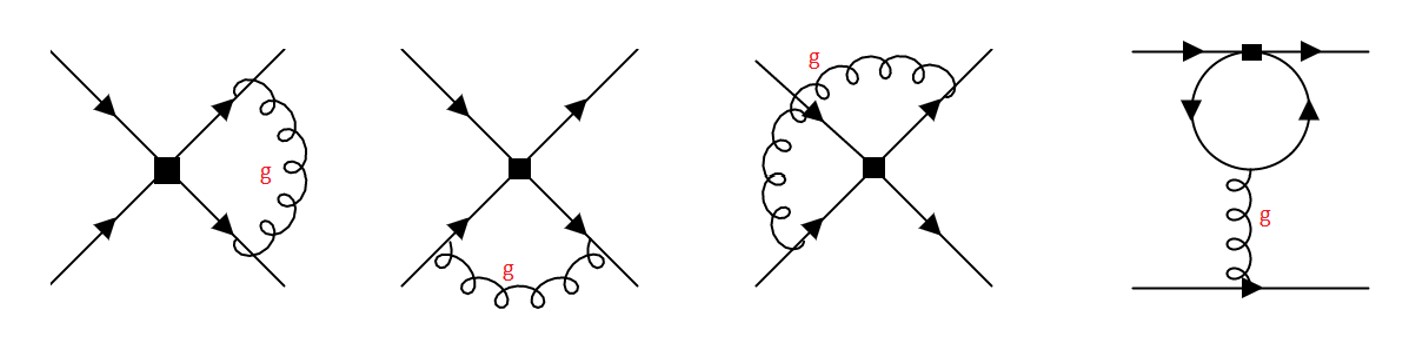}
    \caption{QCD corrections to a four-quark 
    weak process in 
    ${\cal O}(\alpha_s)$.}
    \label{LO-QCD}
\end{figure}

At 
LO, the QCD corrections to the operators 
we consider 
arise from gluon loops as shown in Fig.~\ref{LO-QCD}. 
In the $Z^{0}$ exchange sector, the following operators mix and form a closed set 
under such corrections:
\begin{equation}\label{z ops}
\begin{split}
   \Theta_1 & = [(\bar{u}u)_V+(\bar{d}d)_V+(\bar{s}s)_V]^{\alpha\alpha}[(\bar{u}u)_A-(\bar{d}d)_A-(\bar{s}s)_A]^{\beta\beta}\\
   \Theta_2 & = [(\bar{u}u)_V+(\bar{d}d)_V+(\bar{s}s)_V]^{\alpha\beta}[(\bar{u}u)_A-(\bar{d}d)_A-(\bar{s}s)_A]^{\beta\alpha}\\
   \Theta_3 & = [(\bar{u}u)_A+(\bar{d}d)_A+(\bar{s}s)_A]^{\alpha\alpha}[(\bar{u}u)_V-(\bar{d}d)_V-(\bar{s}s)_V]^{\beta\beta}\\
   \Theta_4 & = [(\bar{u}u)_A+(\bar{d}d)_A+(\bar{s}s)_A]^{\alpha\beta}[(\bar{u}u)_V-(\bar{d}d)_V-(\bar{s}s)_V]^{\beta\alpha}\\
  \Theta_5 & = [(\bar{u}u)_V-(\bar{d}d)_V-(\bar{s}s)_V]^{\alpha\alpha}[(\bar{u}u)_A-(\bar{d}d)_A-(\bar{s}s)_A]^{\beta\beta}\\
  \Theta_6 & = [(\bar{u}u)_V-(\bar{d}d)_V-(\bar{s}s)_V]^{\alpha\beta}[(\bar{u}u)_A-(\bar{d}d)_A-(\bar{s}s)_A]^{\beta\alpha}\\
  \Theta_7 & = [(\bar{u}u)_A+(\bar{d}d)_A+(\bar{s}s)_A]^{\alpha\alpha}[(\bar{u}u)_V+(\bar{d}d)_V+(\bar{s}s)_V]^{\beta\beta}\\
  \Theta_8 & = [(\bar{u}u)_A+(\bar{d}d)_A+(\bar{s}s)_A]^{\alpha\beta}[(\bar{u}u)_V+(\bar{d}d)_V+(\bar{s}s)_V]^{\beta\alpha} \,.
\end{split}
\end{equation}
The extension to include heavier quarks is made possible by the structure shared by \textit{u-like} and \textit{d-like} quarks. For example, once we include all five flavors, 
$\Theta_1$ becomes
\begin{equation}
\Theta_1= [(\bar{u}u)_V+(\bar{c}c)_V+(\bar{d}d)_V+(\bar{s}s)_V+(\bar{b}b)_V]^{\alpha\alpha}[(\bar{u}u)_A+(\bar{c}c)_A-(\bar{d}d)_A-(\bar{s}s)_A-(\bar{b}b)_A]^{\beta\beta} \,.
\end{equation}
Next, using the results from Ref.~\cite{Miller1983anomalousdim} we calculate the anomalous dimension matrix which represents the QCD mixing of the above operators. This serves as a necessary ingredient in performing a RG analysis:
\begin{equation}
\begin{split}
    \gamma_Z(\mu) &=-\frac{g_s^2}{8\pi^2}\begin{pmatrix}
    \frac{2}{9} & \frac{-2}{3} & 1 &-3 & 0 & 0 & 0 & 0\\
    -\frac{3}{2}+\frac{2}{9}n_f & \frac{9}{2}-\frac{2}{3}n_f & \frac{-3}{2} & \frac{-7}{2} & 0 & 0 & 0 & 0\\
    \frac{11}{9} & \frac{-11}{3} & 0 & 0 & 0 & 0 & 0 & 0\\
    \frac{-3}{2} & \frac{-7}{2} & \frac{-3}{2} & \frac{9}{2} & 0 & 0 & -\frac{2}{9}n_Q &\frac{2}{3}n_Q\\
    0 & 0 & 0 & 0 & 1 & -3 & \frac{2}{9} & -\frac{2}{3}\\
    -\frac{2}{9}n_Q & \frac{2}{3}n_Q & 0 & 0 & -3 & 1 & 0 & 0\\
    0 & 0 & 0 & 0 & 0 & 0 & \frac{11}{9} & -\frac{11}{3}\\
    0 & 0 & 0 & 0 & 0 & 0 & \frac{2n_f}{9}-3 & 1-\frac{2n_f}{3} \\ 
    \end{pmatrix} \,,
    \end{split}
\end{equation}
where $n_f$ is the number of dynamical quarks at the considered energy scale and $n_Q=n_d - n_u$, the difference in open $d$-like and $u$-like flavors. 

For the $W$ sector, we first consider the \textit{u-d} current operators that generate and mix in a closed set with
\begin{equation}\label{w ops}
\begin{split}
    &\Theta_9 = (\bar{u}d)_V^{\alpha \alpha}(\bar{d}u)_A^{\beta \beta}+(\bar{d}u)_V^{\alpha \alpha}(\bar{u}d)_A^{\beta \beta}\\
   &\Theta_{10} = (\bar{u}d)_V^{\alpha \beta}(\bar{d}u)_A^{\beta \alpha}+(\bar{d}u)_V^{\alpha \beta}(\bar{u}d)_A^{\beta \alpha}\\
   &\Theta_{ud}^{p_1}  = [(\bar{u}u)_V+(\bar{d}d)_V]^{\alpha\alpha}\sum_{q=u,d,s}(\Bar{q}q)_A^{\beta\beta}+[(\bar{u}u)_A+(\bar{d}d)_A]^{\alpha\alpha}\sum_{q=u,d,s}(\Bar{q}q)_V^{\beta\beta}\\
  &\Theta_{ud}^{p_2}  = [(\bar{u}u)_V+(\bar{d}d)_V]^{\alpha\beta}\sum_{q=u,d,s}(\Bar{q}q)_A^{\beta\alpha}+[(\bar{u}u)_A+(\bar{d}d)_A]^{\alpha\beta}\sum_{q=u,d,s}(\Bar{q}q)_V^{\beta\alpha}\\
  &\Theta_{ud}^{p_3}  = [(\bar{u}u)_A+(\bar{d}d)_A]^{\alpha\alpha}\sum_{q=u,d,s}(\Bar{q}q)_V^{\beta\beta}-[(\bar{u}u)_V+(\bar{d}d)_V]^{\alpha\alpha}\sum_{q=u,d,s}(\Bar{q}q)_A^{\beta\beta}\\
  &\Theta_{ud}^{p_4}  = [(\bar{u}u)_A+(\bar{d}d)_A]^{\alpha\beta}\sum_{q=u,d,s}(\Bar{q}q)_V^{\beta\alpha}-[(\bar{u}u)_V+(\bar{d}d)_V]^{\alpha\beta}\sum_{q=u,d,s}(\Bar{q}q)_A^{\beta\alpha}\\
  &\Theta_{ud}^{p_5}=\sum_{q'=u,d,s}\sum_{q=u,d,s}(\Bar{q}q)_A^{\alpha\alpha}(\Bar{q'}q')_V^{\beta\beta}\\
  &\Theta_{ud}^{p_6}=\sum_{q'=u,d,s}\sum_{q=u,d,s}(\Bar{q}q)_A^{\alpha\beta}(\Bar{q'}q')_V^{\beta\alpha}\\ 
\end{split}
\end{equation}
Here, the $\Theta^{p_i}_{ud}$ operators are generated by penguin insertions. 
Noting Ref.~\cite{Miller1983anomalousdim}, 
the anomalous dimension matrix for the above operators is ultimately found to be 
\begin{equation}\label{w gammabig}
\begin{split}
    \gamma^{ud}_W(\mu) &=-\frac{g_s^2}{8\pi^2}
    \begin{pmatrix}
        1 & -3 & 0 & 0 & 0 & 0 & 0 & 0 \\
        -3 & 1 & \frac{1}{9} & \frac{-1}{3} & \frac{1}{9} & \frac{-1}{3} & 0 & 0\\
        0 & 0 & \frac{11}{9} & \frac{-11}{3} & \frac{2}{9} & \frac{-2}{3} & 0 & 0\\
         0 & 0 & -3+\frac{n_f}{9} & 1-\frac{n_f}{3} & \frac{n_f}{9} & \frac{-n_f}{3} & \frac{2}{9} & \frac{-2}{3}\\
          0 & 0 & 0 & 0 & -1 & 3 & 0 & 0\\
           0 & 0 & \frac{n_f}{9} & \frac{-n_f}{3} & \frac{n_f}{9} & \frac{n_f}{3}-8 & \frac{-2}{9} & \frac{2}{3}\\
            0 & 0 & 0 & 0 & 0 & 0 & \frac{11}{9} & \frac{-11}{3}\\
             0 & 0 & 0 & 0 & 0 & 0 & \frac{2n_f}{9}-3 & 1-\frac{2n_f}{3}\\
    \end{pmatrix} \,.
    \end{split}
\end{equation}
The operator set from the \textit{u-s} current 
and corresponding anomalous dimension matrix can be found by replacing $d\rightarrow s$. Above the charm-quark mass threshold, the effects of penguin-generated $\Theta^{p_i}$ operators of $u$-like quarks 
cancel each other, as in the $\Delta S=1$ sector~\cite{Buchalla:1995vs},
due to the Glashow–Iliopoulos–Maiani mechanism, though the analysis below the charm mass scale is richer in the flavor-diagonal case. 
Nevertheless, the effect of renormalization group flow 
below the 
charm-quark mass scale to the hadronic scale 
of $\mu=2\, \rm GeV$ is negligible compared 
to the LO evolution of 
the rest of the operators. 
Thus, along with the \textit{u-s} current operators,
\begin{equation}\label{w ops 2nd set}
\begin{split}
    \Theta_{11} &= (\bar{u}s)_V^{\alpha \alpha}(\bar{s}u)_A^{\beta \beta}+(\bar{s}u)_V^{\alpha \alpha}(\bar{u}s)_A^{\beta \beta}\\
    \Theta_{12} &= (\bar{u}s)_V^{\alpha \beta}(\bar{s}u)_A^{\beta \alpha}+(\bar{s}u)_V^{\alpha \beta}(\bar{u}s)_A^{\beta \alpha} \,,
\end{split}
\end{equation}
we simplify the anomalous dimension matrix for the $W$ sector as 
\begin{equation}\label{w gamma}
\begin{split}
    \gamma_W(\mu) &=-\frac{g_s^2}{8\pi^2}
    \begin{pmatrix}
        1 & -3 & 0 & 0\\
        -3 & 1 & 0 & 0\\
        0 & 0 & 1 & -3\\
        0 & 0 & -3 & 1
    \end{pmatrix} \,.
    \end{split}
\end{equation}
for the set $\Theta_{9}, \Theta_{10}, \Theta_{11}$, and $\Theta_{12}$. We now turn to our numerical analysis. 



\section{Renormalization group flow}
We start by writing the PV Hamiltonian in Eq.~(\ref{PVH}) compactly as
\begin{equation}
        \mathcal{H}_{\rm eff}^{\rm PV}(\mu) = \frac{G_Fs_w^2}{3\sqrt{2}}\sum_{i=1}^{12} C_i(\mu)\Theta_i \,.
\end{equation}
The Wilson coefficients $C_i$ flow between different energy scales according to the equation
\begin{equation}
        \Vec{C}(\mu) = \exp\left[\int_{g_s(M_W)}^{g_s(\mu)}dg \frac{\gamma^T(\mu)}{\beta(g_s)}\right]\Vec{C}(M_W) \,.
\end{equation}
Here the $\gamma$ matrices from the $W$ and $Z$ sectors are combined as $\gamma = \gamma_z \oplus \gamma_w$ and the QCD $\beta$ function is 
\begin{equation}
        \beta(g_s) = -\frac{g_s^3}{48\pi^2}(33-2n_f) \,.
\end{equation}
As our work is limited to a LO analysis, 
we use the one-loop expression for 
the strong-coupling parameter $\alpha_s(\mu)=g_s^2/4\pi$:
\begin{equation}
    \alpha_s(\mu) = \frac{4\pi}{\beta_0\,  {\rm ln} (\mu^2/\Lambda^2)} \quad {\rm with} \quad \beta_0 = \frac{1}{3}(33-2n_f) \,,
\end{equation}
but we have used the two-loop expression for $\alpha_s(\mu)$ to set the QCD scale parameters in different energy ranges. With the input value at $Z^0$ mass 
scale, $\alpha_s(M_z) = 0.117$~\cite{Zyla:2020zbs}, 
the criterion of continuity across the 
heavy quark flavor thresholds
leads to the following scale parameters: for five-flavor QCD, $\Lambda_{5}=0.214 \,\rm GeV$, for four-flavor QCD, $\Lambda_{4}= 0.307 \,\rm GeV$, and for three-flavor QCD, $\Lambda_{3}=0.352 \,\rm GeV$, as this improves the LO analysis considerably. 
The resulting strong interaction 
strength ratios at LO [next-to-leading order (NLO)]
are
\begin{eqnarray}
  &&\frac{\alpha_s(M_b=4.18\,\rm GeV)}{\alpha_s(M_W=80.379\,\rm GeV)}= 2.09\, [1.86]; \, \frac{\alpha_s(M_c=1.27\,\rm GeV)}{\alpha_s(M_b)}= 1.88\, [1.75]; \nonumber\\
  &&\frac{\alpha_s(2\,\rm GeV)}{\alpha_s(M_c)}=0.74\, [0.75]\,,
\end{eqnarray}
whereas after matching at next-to-next-to-leading order (NNLO), employing the convenient RunDec package~\cite{Chetyrkin:2000yt},
the ratios at that order become 1.85, 1.68, and 0.77.
Performing the RG flow from the energy scale $M_W$ to the hadronic scale of $2\,{\rm GeV}$, 
$\Vec{C}(M_W) = (1, 0, 0, 0, -3.49, 0, 0, 0, -13.0\cos^2\!\theta_c, 0, -13.0\sin^2\!\theta_c, 0 )$
evolves to 
\begin{equation} 
        \!\!\!\!\!\!\!\!\!\!\Vec{C}(2 \,\rm GeV)\! = \!\begin{pmatrix} 
            1.09 &[1.17 \dots 1.06] [1.08 \dots 1.04] &[1.07] [1.06]\\
            0.018 &[0.014 \dots 0.021] [0.033 \dots 0.006] &[-0.006] [-0.006]\\
            0.199 &[0.321\dots 0.133] [0.193 \dots 0.127] &[0.158] [0.153]\\
            -0.583 &[-0.990 \dots -0.385] [-0.571 \dots -0.374] &[-0.460] [-0.456]\\
            -4.36  &[-4.99 \dots -4.05] [-4.34 \dots -4.03] &[-4.16] [-4.14]\\
            1.72 &[2.63 \dots 1.19] [1.67 \dots 1.16]  &[1.40] [1.36]\\
            -0.170 &[-0.288 \dots -0.110] [-0.165 \dots -0.105] &[-0.134] [-0.129]\\
            0.332 &[0.496 \dots 0.235] [0.322 \dots 0.225] &[0.275] [0.268]\\
            -16.2 &[-18.6 \dots -15.0] [-16.1 \dots -15.0] &[-15.48] [-15.4] \\
            6.38 &[9.76 \dots 4.44] [6.22 \dots 4.30]  &[5.19] [5.05] \\
            -16.2 &[-18.6 \dots -15.0] [-16.1 \dots -15.0] &[-15.48] [-15.4] \\
            6.38  &[9.76 \dots 4.44] [6.22 \dots 4.30] &[5.19] [5.05]
        \end{pmatrix} \,,
        \label{eq:fullheff}
\end{equation}
where the last four entries should be multiplied by factors of 
 $\cos^2\!\theta_c,  \cos^2\!\theta_c,  \sin^2\!\theta_c$, and $\sin^2\!\theta_c$, respectively. 
Our primary result 
is given by the leftmost column of numbers, and the coefficients have been 
simplified with the substitution $s_W^2=0.231$. The other columns illustrate the uncertainties in our computation. 
It should be noted that we perform RG-flow below $2\,\rm GeV$ to integrate out the \textit{charm} quark and then run upwards 
so as to work with a $N_f=2+1$ theory at $2\,\rm GeV$. An alternate would be to evolve 
to $2\, \rm GeV$ with a $N_f=2+1+1$ theory and consider
the $u,d,s$ contributions to $H_{\rm eff}$ only. 
The resulting 
Wilson coefficients 
are very similar in the two approaches. 
In the 
central column of Eq.~(\ref{eq:fullheff}), the left set shows the ranges of 
Wilson coefficients 
that result 
in the $N_f=2+1$ theory for renormalization scales of $\mu=1-4 \,{\rm GeV}$ and the right set 
shows them in the $N_f=2+1 +1$ theory with $\mu=2-4 \,{\rm GeV}$. In the rightmost column, we show the 
Wilson coefficients 
if the $\alpha_s$ running and matching is computed at NLO (left) and NNLO (right).

We conclude this section by comparing our results
with those of Ref.~\cite{Desplanques:1979hn}, in which the QCD evolution effects were estimated 
via a phenomenological enhancement factor, $K$:
\begin{equation}
    K = 1+\frac{g^2(\mu^2)}{16\pi^2}\Bigg( 11-\frac{2}{3}n_f \Bigg)\,\ln\Bigg( \frac{M^2_W}{\mu^2}\Bigg) \,,
    \label{eq:Kdef}
\end{equation}
where $\mu$ is any energy scale below $M_W$, $n_f$ is the number of dynamical quarks at scale $\mu$, and operator mixing is not
included. 
The 
Wilson coefficients 
at hadronic scales corresponding to $K=4$ ($\mu\approx 1.6 \,\rm GeV $), 
after adjusting for an overall sign difference due to differing sign
conventions, are found to be
$C_1^{\rm DDH}=1.15$, $C_2^{\rm DDH}=0$, $C_3^{\rm DDH} = 0$, $C_4^{\rm DDH}=-0.39$, $C_5^{\rm DDH}=-3.95$,$C_6^{\rm DDH}=1.08$, $C_7^{\rm DDH}=-0.44$, $C_8^{\rm DDH}=1.04$, $C_9^{\rm DDH}=-14.67 \cos^2\!\theta_c$, $C_{10}^{\rm DDH}=4.02 \cos^2\!\theta_c$, $C_{11}^{\rm DDH}=-14.67 \sin^2\!\theta_c$ and $C_{12}^{\rm DDH}=4.02 \sin^2\!\theta_c$.
Comparing to Eq.~(\ref{eq:fullheff}), we observe that the QCD operator mixing and flavor threshold effects necessary
for a complete calculation play an important numerical role, with earlier
results~\cite{Desplanques:1979hn} 
falling outside the range possible
through the consideration of scale variation and higher-order effects.

\section{Isosector extractions}
With the full PV effective Hamiltonian in hand, we extract the contributions from the individual isosectors. For example, 
for the isovector sector:
\begin{equation}
        \mathcal{H}_{\rm eff}^{\rm PV}(\mu) = \frac{G_Fs_w^2}{3\sqrt{2}}\sum_{i=1}^{12} C_i(\mu) \Theta_i \longrightarrow \mathcal{H}^{I=1}_{\rm eff}(\mu) = \frac{G_Fs_w^2}{3\sqrt{2}}\sum_{i=1}^{12} C^{I=1}_i(\mu) \Theta_i^{I=1}
\end{equation}
Considering the 
operators in Eqs.(\ref{z ops}, \ref{w ops}, \ref{w ops 2nd set}) we see
$\Theta_{1-4,11,12}$ contribute one operator each: $C_j\Theta_j \rightarrow C^{I=1}_j\Theta^{I=1}_j$ with $C_j=C^{I=1}_j$.
Operators $\Theta_5$ and $\Theta_6$ contribute two operators each: 
$C_5\Theta_5 \rightarrow C^{I=1}_5\Theta_5^{I=1} + C^{I=1}_7\Theta_7^{I=1}$ with $C^{I=1}_5 = C^{I=1}_7 = -C_5$ 
and $C_6\Theta_6 \rightarrow C^{I=1}_6\Theta_6^{I=1} + C^{I=1}_8\Theta_8^{I=1}$ 
with $C^{I=1}_6 = C^{I=1}_8 = -C_6$.  
Operators $\Theta_{7-10}$ have no contributions to the isovector sector. Thus the extracted isovector operator set is 
\begin{equation}\label{isovec op set}
    \begin{split}
        \Theta_1&^{I=1}= [(\bar{u}u)_V+(\bar{d}d)_V+(\bar{s}s)_V]^{\alpha\alpha}[(\bar{u}u)_A-(\bar{d}d)_A]^{\beta\beta}\\
        \Theta_2&^{I=1}= [(\bar{u}u)_V+(\bar{d}d)_V+(\bar{s}s)_V]^{\alpha\beta}[(\bar{u}u)_A-(\bar{d}d)_A]^{\beta\alpha}\\
        \Theta_3&^{I=1}= [(\bar{u}u)_A+(\bar{d}d)_A+(\bar{s}s)_A]^{\alpha\alpha}[(\bar{u}u)_V-(\bar{d}d)_V]^{\beta\beta}\\
        \Theta_4&^{I=1}= [(\bar{u}u)_A+(\bar{d}d)_A+(\bar{s}s)_A]^{\alpha\beta}[(\bar{u}u)_V-(\bar{d}d)_V]^{\beta\alpha}\\
        \Theta_5&^{I=1}= (\bar{s}s)_V^{\alpha\alpha}[(\bar{u}u)_A-(\bar{d}d)_A]^{\beta\beta}\\
        \Theta_6&^{I=1}= (\bar{s}s)_V^{\alpha\beta}[(\bar{u}u)_A-(\bar{d}d)_A]^{\beta\alpha}\\
        \Theta_7&^{I=1}= (\bar{s}s)_A^{\alpha\alpha}[(\bar{u}u)_V-(\bar{d}d)_V]^{\beta\beta}\\
        \Theta_8&^{I=1}= (\bar{s}s)_A^{\alpha\beta}[(\bar{u}u)_V-(\bar{d}d)_V]^{\beta\alpha}\\
        \Theta_9&^{I=1}= (\bar{u}s)_V^{\alpha\alpha}(\bar{s}u)_A^{\beta\beta} + (\bar{s}u)_V^{\alpha\alpha}(\bar{u}s)_A^{\beta\beta}\\
        \Theta_{10}&^{I=1}= (\bar{u}s)_V^{\alpha\beta}(\bar{s}u)_A^{\beta\alpha} + (\bar{s}u)_V^{\alpha\beta}(\bar{u}s)_A^{\beta\alpha}\\
        \end{split} \,,
\end{equation}
and 
the extracted isovector 
Wilson coefficients 
at high and low energies 
are: $\Vec{C}^{I=1}(M_W) = (1, 0, 0, 0, 3.49, 0, 3.49, 0, -13.0\cos^2\!\theta_c, 0)$ and
\begin{equation}\label{vec wil coef}
        \!\!\!\!\!\!\!\!\!\!\Vec{C}^{I=1}(2 \,\rm GeV)\! = \!\begin{pmatrix} 
            1.09 &[1.17 \dots 1.06] [1.08 \dots 1.04] &[1.07] [1.06]\\
            0.018 &[0.014 \dots 0.021] [0.033 \dots 0.006] &[-0.006] [-0.006]\\
            0.199 &[0.321\dots 0.133] [0.193 \dots 0.127] &[0.158] [0.153]\\
            -0.583 &[-0.990 \dots -0.385] [-0.571 \dots -0.374] &[-0.460] [-0.456]\\
            4.36  &[4.99 \dots 4.05] [4.34 \dots 4.03] &[4.16] [4.14]\\
            -1.72 &[-2.63 \dots -1.19] [-1.67 \dots -1.16]  &[-1.40] [-1.36]\\
            4.36  &[4.99 \dots 4.05] [4.34 \dots 4.03] &[4.16] [4.14]\\
            -1.72 &[-2.63 \dots -1.19] [-1.67 \dots -1.16]  &[-1.40] [-1.36]\\
            -16.2 &[-18.6 \dots -15.0] [-16.1 \dots -15.0] &[-15.48] [-15.4] \\
            6.38 &[9.76 \dots 4.44] [6.22 \dots 4.30]  &[5.19] [5.05] \\
        \end{pmatrix} \,,
        \end{equation}
where the last two entries should be multiplied by a factor $\sin^2\!\theta_c$ and 
the error estimates are defined as in Eq.~(\ref{eq:fullheff}). 
Alternatively, an
isovector RG analysis can be performed directly to get $\Vec{C}^{I=1}(2\,\rm GeV)$ from $\Vec{C}^{I=1}(M_W)$ using the anomalous dimension matrix corresponding to the operator set in Eq.(\ref{isovec op set}):
\begin{equation}
\begin{split}
    \gamma_Z^{I=1} &=-\frac{g_s}{8\pi^2}\begin{pmatrix}
    \frac{2}{9} & \frac{-2}{3} & 1 &-3 & 0 & 0 & 0 & 0\\
    -\frac{3}{2}+\frac{2}{9}n_f & \frac{9}{2}-\frac{2}{3}n_f & \frac{-3}{2} & \frac{-7}{2} & 0 & 0 & 0 & 0\\
    \frac{11}{9} & \frac{-11}{3} & 0 & 0 & 0 & 0 & 0 & 0\\
    \frac{-3}{2} & \frac{-7}{2} & \frac{-3}{2} & \frac{9}{2} & 0 & 0 & 0 & 0\\
    0 & 0 & 0 & 0 & 0 & 0 & 1 & -3\\
    \frac{2}{9}n_Q & -\frac{2}{3}n_Q & 0 & 0 & \frac{-3}{2} & \frac{9}{2} & \frac{-3}{2} & \frac{-7}{2}\\
    0 & 0 & 0 & 0 & 1 & -3 & 0 & 0\\
    0 & 0 & 0 & 0 & \frac{-3}{2} & \frac{-7}{2} & \frac{-3}{2} & \frac{9}{2}\\
    \end{pmatrix}
    \end{split}
\end{equation}
The corresponding $W$ sector matrix can be easily obtained 
from Eq.~(\ref{w gamma}). 
The set of 
Wilson coefficients 
obtained via RG flow exactly matches the results in 
Eq.~(\ref{vec wil coef}), in agreement with Fig. (\ref{Fig:rectangle}).  A purely 
isovector $Z^{0}$ sector RG analysis was performed in Ref.~\cite{Dai:1991bx}. 
Our 
results are in agreement when the same inputs are used.

We can make a similar analysis 
in the $I=0 \oplus 2$ sector. The corresponding operators are
\begin{equation}\label{isoeven ops}
    \begin{split}
        \Theta_1&^{I=0 \oplus 2}= [(\bar{u}u)_V+(\bar{d}d)_V+(\bar{s}s)_V]^{\alpha\alpha}[(\bar{s}s)_A]^{\beta\beta}\\
        \Theta_2&^{I=0 \oplus 2}= [(\bar{u}u)_V+(\bar{d}d)_V+(\bar{s}s)_V]^{\alpha\beta}[(\bar{s}s)_A]^{\beta\alpha}\\
         \Theta_3&^{I=0 \oplus 2}= [(\bar{u}u)_A+(\bar{d}d)_A+(\bar{s}s)_A]^{\alpha\alpha}[(\bar{s}s)_V]^{\beta\beta}\\
        \Theta_4&^{I=0 \oplus 2}= [(\bar{u}u)_A+(\bar{d}d)_A+(\bar{s}s)_A]^{\alpha\beta}[(\bar{s}s)_V]^{\beta\alpha}\\
        \Theta_5&^{I=0 \oplus 2}= [(\bar{u}u)_V-(\bar{d}d)_V]^{\alpha\alpha}[(\bar{u}u)_A-(\bar{d}d)_A]^{\beta\beta}+(\bar{s}s)_V^{\alpha\alpha}(\bar{s}s)_A^{\beta\beta}\\
        \Theta_6&^{I=0 \oplus 2}= [(\bar{u}u)_V-(\bar{d}d)_V]^{\alpha\beta}[(\bar{u}u)_A-(\bar{d}d)_A]^{\beta\alpha}+(\bar{s}s)_V^{\alpha\beta}(\bar{s}s)_A^{\beta\alpha}\\
       \Theta_7&^{I=0 \oplus 2}= [(\bar{u}u)_V+(\bar{d}d)_V+(\bar{s}s)_V]^{\alpha\alpha}[(\bar{u}u)_A+(\bar{d}d)_A+(\bar{s}s)_A]^{\beta\beta}\\
       \Theta_8&^{I=0 \oplus 2}= [(\bar{u}u)_A+(\bar{d}d)_A+(\bar{s}s)_A]^{\alpha\beta}[(\bar{u}u)_V+(\bar{d}d)_V+(\bar{s}s)_V]^{\beta\alpha}\\
        \Theta_9&^{I=0 \oplus 2}= (\bar{u}d)_V^{\alpha\alpha}(\bar{d}u)_A^{\beta\beta} + (\bar{d}u)_V^{\alpha\alpha}(\bar{u}d)_A^{\beta\beta}\\
        \Theta_{10}&^{I=0 \oplus 2}= (\bar{u}d)_V^{\alpha\beta}(\bar{d}u)_A^{\beta\alpha} + (\bar{d}u)_V^{\alpha\beta}(\bar{u}d)_A^{\beta\alpha}\\
        \end{split} \,,
\end{equation}
and the 
extracted 
Wilson coefficients 
for
the $I=0 \oplus 2$ sector 
at high and low energies are: 
$\Vec{C}^{I=0\oplus2}(M_W) = (-1, 0, 0, 0, -3.49, 0, 0, 0, -13.0\cos^2\!\theta_c, 0)$ and  
\begin{equation} 
        \!\!\!\!\!\!\!\!\!\!\Vec{C}^{ I=0\oplus2}(2\,\rm GeV)\! = \!\begin{pmatrix} 
            -1.09 &[-1.17 \dots -1.06] [-1.08 \dots -1.04] &[-1.07] [-1.06]\\
            -0.018 &[-0.014 \dots -0.021] [-0.033 \dots -0.006] &[0.006] [0.006]\\
            -0.199 &[-0.321\dots -0.133] [-0.193 \dots -0.127] &[-0.158] [-0.153]\\
            0.583 &[0.990 \dots 0.385] [0.571 \dots 0.374] &[0.460] [0.456]\\
            -4.36  &[-4.99 \dots -4.05] [-4.34 \dots -4.03] &[-4.16] [-4.14]\\
            1.72 &[2.63 \dots 1.19] [1.67 \dots 1.16]  &[1.40] [1.36]\\
            -0.170 &[-0.288 \dots -0.110] [-0.165 \dots -0.105] &[-0.134] [-0.129]\\
            0.332 &[0.496 \dots 0.235] [0.322 \dots 0.225] &[0.275] [0.268]\\
            -16.2 &[-18.6 \dots -15.0] [-16.1 \dots -15.0] &[-15.48] [-15.4] \\
            6.38 &[9.76 \dots 4.44] [6.22 \dots 4.30]  &[5.19] [5.05] \\
        \end{pmatrix}
        \label{sca ten wil coef}
\end{equation}
where the last two entries should be multiplied by a factor $\cos^2\!\theta_c$ and 
the error estimates are defined as in Eq.~(\ref{eq:fullheff}). 
If one wishes to perform a 
RG analysis of the isoeven sectors to obtain $\Vec{C}^{I=0\oplus2}(2\,\rm GeV)$ from $\Vec{C}^{I=0\oplus2}(M_W)$, the anomalous dimension matrix for the operator set in Eq.(\ref{isoeven ops}) is
\begin{equation}
\begin{split}
    \gamma^{I=0\oplus2}_Z &=-\frac{g_s}{8\pi^2}\begin{pmatrix}
    \frac{2}{9} & \frac{-2}{3} & 1 &-3 & 0 & 0 & 0 & 0\\
    -\frac{3}{2}+\frac{2}{9}n_f & \frac{9}{2}-\frac{2}{3}n_f & \frac{-3}{2} & \frac{-7}{2} & 0 & 0 & 0 & 0\\
    \frac{11}{9} & \frac{-11}{3} & 0 & 0 & 0 & 0 & 0 & 0\\
    \frac{-3}{2} & \frac{-7}{2} & \frac{-3}{2} & \frac{9}{2} & 0 & 0 & \frac{2}{9}n_Q &-\frac{2}{3}n_Q\\
    0 & 0 & 0 & 0 & 1 & -3 & \frac{2}{9} & -\frac{2}{3}\\
    \frac{2}{9}n_Q &-\frac{2}{3}n_Q & 0 & 0 & -3 & 1 & 0 & 0\\
    0 & 0 & 0 & 0 & 0 & 0 & \frac{11}{9} & -\frac{11}{3}\\
    0 & 0 & 0 & 0 & 0 & 0 & \frac{2n_f}{9}-3 & 1-\frac{2n_f}{3}\\
    \end{pmatrix}
    \end{split} \,,
\end{equation}
and the $W$ sector matrix can be easily obtained from Eq.~(\ref{w gamma}). Again, the set of
Wilson coefficients 
obtained via RG flow exactly matches the extracted coefficients sets in 
Eq.~(\ref{sca ten wil coef}), 
in agreement with Fig. (\ref{Fig:rectangle}).

\section{Estimation of the parity-violating pion-NN coupling constant}
We can now use our effective Hamiltonian to compute the 
parity-violating meson-NN
coupling constants of isospin $I$, $h_M^I$, that appear in the 
phenomenological Hamiltonian $\mathcal{H}_{\rm DDH}$~\cite{Desplanques:1979hn}, to compare and contrast with earlier estimates. 
For example, the one-pion contribution to hadronic parity violation stems from
\begin{equation}
    \mathcal{H}^{\pi}_{\rm DDH} = i h^1_{\pi}(\pi^{+}\bar{p}n-\pi^{-}\bar{n}p) \,,
\end{equation}
implying that $h_\pi^1$
can be determined via 
\begin{equation}
    -i h_{\pi}^1 \bar{u}_n u_p = \bra{n\pi^{+}}\mathcal{H}_{\rm eff}^{I=1}\ket{p}\,,
    \label{pimatch}
\end{equation}
where $u_N$ with $N\in p,n$ is a Dirac spinor. Using the factorization approximation~\cite{Michel:1964zz,Bauer:1986bm} 
on the matrix element of Eq.~(\ref{pimatch}), 
we have
\begin{equation}
     h_{\pi}^1 \bar{u}_n u_p=\frac{2G_Fs_w^2}{3\sqrt{2}}\left(\frac{C_{1}^{I=1}}{N_c}+C_{2}^{I=1}-\frac{C_{3}^{I=1}}{N_c}-C_{4}^{I=1}\right)\frac{m_{\pi}^2 f_{\pi}}{(m_u+m_d)}
          \bra{n}\bar{d}u\ket{p}\,,
     \label{evalhpi1}
\end{equation}
where $N_c=3$
and we have simplified our result using the quark-field equations of motion
and $\langle 0 | (\bar d u)_A (0) |  \pi^+ (p) \rangle \equiv i p^\mu f_\pi$ to find 
\begin{equation}
    \bra{\pi^+} (\bar{u}   \gamma_{5} d) \ket{0} = \frac{m_{\pi}^2 f_{\pi}}{i(m_u+m_d)} \,.
\end{equation}
For the numerical evaluation of Eq.~(\ref{evalhpi1}) we use $m_\pi=135\,\rm MeV$ and the 
charged-pion decay constant $f_\pi=130\,\rm MeV$, with the other inputs coming from $N_f=2+1$ lattice QCD (LQCD) results: the renormalization-group-invariant (RGI) mass $(m_u+m_d) = 2(4.695 (56)_m (54)_\Lambda\,\rm MeV)$ ~\cite{Aoki:2021kgd} and the 
isovector quark scalar charge of the nucleon ($\bra{n}\bar{d}u\ket{p}\equiv g_s^{u-d}\bar{u}_nu_p$) with $g_s^{u-d}=1.06(10)(06)_{sys}$~\cite{Park:2021ypf} at $\mu=2\,{\rm GeV}$. 
Using Eq.~(\ref{vec wil coef}) we find 
\begin{equation}
 h_{\pi}^1  = (3.06 \pm 0.34  + \left(\stackrel{{+1.29}}{{}_{-0.64}}\right) + 0.42 + (1.00))\times 10^{-7} \,,
 \label{eq:hpi1}
\end{equation}
where the error estimates come, respectively, 
from the LQCD inputs we employ and from the 
following systematic effects: the change in the Wilson coefficients over (i) a scale variation 
of $1-4\,{\rm GeV}$ and (ii) higher-order corrections in $\alpha_s$ as per Eq.~(\ref{vec wil coef}), 
and, finally, our estimate of the accuracy of Eq.~(\ref{evalhpi1}) through the contribution 
to it from ${\cal O}(1/N_c)$ terms, which we note in parentheses. We emphasize
that this last could be an underestimate of the uncertainty. 
To compare,  
the DDH ``best value'' of $h_\pi^1=4.6\times 10^{-7}$~\cite{Desplanques:1979hn} neglects operator mixing 
as noted after Eq.(\ref{eq:Kdef}), although that value is driven by their estimate of 
non-factorizable contributions, which may be grossly overestimated~\cite{Dubovik:1986pj}.
A computation with Wilson coefficients compatible with ours yields 
$h_\pi^1 \sim 1.5\times 10^{-7}$ at $\mu=1\,{\rm GeV}$ using the factorization approximation, a 
SU(3)$_f$-based assumption for the nucleon matrix element,
and phenomenological fits for the light quark mass ratios
with $m_s=200\,{\rm MeV}$ --- but $h_\pi^1 \sim 6\times 10^{-7}$ using SU(3)$_f$ transformations and 
hyperon data~\cite{kaplan1993analysis}, trends also noted in two- and three-flavor Skyrme 
models~\cite{Kaiser:1988bt,Meissner:1998pu}, albeit their outcomes are much smaller.
These large variations in $h_\pi^1$ 
are remediated through the use of LQCD for the
nucleon matrix element. 
The experimental result $h_\pi^1 = 2.6(1.2)_{\rm stat}(0.2)_{\rm sys}\times10^{-7}$, determined from the 
parity-violating gamma asymmetry in ${\vec n} p \to d \gamma$~\cite{NPDGamma:2018vhh}, is also 
comparable to its value extracted using chiral EFT~\cite{deVries:2015pza,deVries:2020iea}, and 
we believe 
our improved comparison to it 
affirms the validity of our approach. We refer to Ref.~\cite{Gardner:2022mxf} for 
further discussion and broader examples. 

\section{Summary} 
We have determined the effective 
weak Hamiltonian for parity-violating, $\Delta S=0$
hadronic processes
in the Standard Model at a renormalization scale of $2\, \rm GeV$. 
To do this, we have made a 
complete, LO 
renormalization
group analysis in QCD, starting from just below the $W$ mass
scale, including operator mixing and 
evolution through heavy-flavor thresholds, as well as 
neutral- and charged-current effects, for all possible
isospins ($I=0,1,2$) of the four-quark operators. 
In our analysis we have found it convenient to separate 
the $I=1$ and $I=0\oplus 2$ sectors, and isospin
symmetry allows us to 
recover the same low-energy effective Hamiltonian 
regardless of the order in which we (i) evolve to low-energy
scales or (ii) project on operators with even or odd isospin
--- a test, which we note here for the first time, that 
should prove particularly useful in a future NLO analysis. 
The construction of the 
complete effective weak Hamiltonian at a scale of 
$\mu =2\,\rm GeV$ should support LQCD studies
of two-nucleon matrix elements~\cite{Nicholson:2021zwi}, enabling further
theoretical studies in which the factorization
approximation of the hadronic matrix elements
would finally be no longer necessary. 

\section*{Acknowledgments}
We acknowledge partial support from the U.S. Department of Energy Office
of Nuclear Physics under contract DE-FG02-96ER40989.
We thank the INT for gracious hospitality and 
acknowledge 
lively discussions with the workshop participants 
of ``Hadronic Parity Nonconservation II'' while
this work was being completed.

\bibliography{HPV_refs_bib}

\end{document}